\documentclass[]{aastex62}
\usepackage{natbib}
\usepackage{graphicx}
\usepackage{amsmath}
\usepackage{amssymb}
\usepackage{ctable}
\begin{document}
\title{A tangle of stellar streams in the north Galactic cap}
\author{Jake Weiss}
\affiliation{Department of Physics, Applied Physics and Astronomy, Rensselaer Polytechnic Institute, Troy, NY 12180, USA}
\author{Heidi Jo Newberg}
\affiliation{Department of Physics, Applied Physics and Astronomy, Rensselaer Polytechnic Institute, Troy, NY 12180, USA}
\author{Travis Desell}
\affiliation{Department of Software Engineering, Rochester Institute of Technology, 134 Lomb Memorial Drive, Rochester, NY 14623, USA}
\begin{abstract}

Stellar halo substructures were identified using statistical photometric parallax of blue main sequence turnoff stars from fourteen Sloan Digital Sky Survey stripes in the north Galactic cap.  Four structures are consistent with previous measurements of the Sagittarius dwarf tidal stream: the leading tail, the ``bifurcated" stream, the trailing tail, and Stream C.  The stellar overdensity in Virgo, about 15 kpc from the Sun, could arise from the crossing of the Parallel Stream and a new, candidate stream dubbed the Perpendicular Stream.  The data suggests the presence of a wide stream near NGC 5466, with a distance of 5 to 15 kpc.  Measurements of the flattening of the smooth stellar halo from the fourteen stripes average $q=0.58$, with a dispersion of 0.04.

\end{abstract}
\keywords{catalogs --- Galaxy: halo --- Galaxy: structure --- methods: data analysis --- methods: statistical}

\section{Introduction}

Statistical photometric parallax \citep{newberg2013} is a method for determining the underlying density distribution of a stellar population, using statistical knowledge of the absolute magnitudes of the stars in that population.  This technique is particularly useful for tracing density distributions using main sequence turnoff (MSTO) stars, which vary in intrinsic brightness by about two magnitudes.  MSTO stars from the Sloan Digital Sky Survey \citep[SDSS;][]{SDSSYork} have been used to discover density substructure in the Milky Way's stellar halo \citep{newberg2002}, to create the Field of Streams image \citep{Belokurov2006}, and to trace the stellar halo number density \citep{Juric2008}.  In older stellar populations, MSTO stars are the brightest main sequence stars, which are far more numerous than giant stars.  The method for applying statistical photometric parallax to SDSS MSTO stars to obtain density measurements of halo substructure has been developed in \citet{cole2008, newby2011, newby2013, WeissSubmitted}.

MilkyWay@home \citep{Newberg2014} is a powerful volunteer computing network, delivering $\sim 1$ petaFLOPS of computing power to statistical photometric parallax calculations.  Each node computes the likelihood of one set of density model parameters, given the observed positions and magnitudes of turnoff stars.  The optimization algorithms on MilkyWay@home are described in \citet{desell2007, desell2010, desell2011, desell2017}.  \citet{WeissSubmitted} demonstrated that MilkyWay@home is capable of recovering the parameters of three tidal streams and a smooth background distribution in a single $2.5^\circ$-wide SDSS stripe of data.  Here, we use MilkyWay@home to fit stripes 10 through 23 in the SDSS north Galactic cap.

\section{Data Selection}

We select our MSTO stars from SDSS Data Release 7 \citep[DR7;][]{DR7Abazajian2009}, with the criteria: $g_0>16$, $0.1<(g - r)_0<0.3$, $(u - g)_0>0.4$, and EDGE and SATURATED flags not set \citep{NewbergYanny2006, newberg2002}.  We use the subscript ``0" to indicate that the magnitudes are reddening corrected using the \cite{SFD1998} dust maps.  The effects of reddening are small since we select stars with $b>30^{\circ}$.  The $(g - r)_0$ cut picks out bluer MSTO stars, while avoiding redder thick disk MSTO stars. The $(u - g)_0>0.4$ cut eliminates quasars. The length of the stripes is limited to minimize contamination from low latitude substructure.  We also removed a region of the sky around each of the globular clusters: NGC 4147, NGC 5024, NGC 5053, NGC 5272, and NGC 5466.

An SDSS ``wedge" includes a volume defined by the angular limits of a stripe and the heliocentric distance to the most distant object in the dataset.  Each SDSS stripe is $2.5^\circ$ wide, and typically $140^{\circ}$ long.  Most MSTO stars are within 45 kpc of the Sun, though intrinsically brighter MSTO stars can be observed to about twice this distance.  Since the density varies only a small amount in the narrow direction, we often depict the stellar density in each stripe using a polar plot with the radius proportional to the distance and the angle given by $\mu$, the angular distance along the stripe in SDSS Great Circle coordinates.  Sample wedge plots are shown in the leftmost panels of Figure \ref{FullSkyResiduals}.

\section{Model Overview}

We model the Milky Way halo on a stripe by stripe basis with a Hernquist profile \citep{hernquist}, an exponential thick disk, and three tidal streams.  A detailed description of our density model and fitting technique can be found in \citet{WeissSubmitted}.

The scale radius of the Hernquist profile is fixed at 12 kpc.  We fit a flattening parameter, $q$, and weighting factor, $\varepsilon_{sph}$.  The flattening parameter allows us to fit an oblate ($q<1$) or prolate ($q>1$) spheroid. The weighting factor indicates the fraction of the smooth background (Hernquist plus disk) stars assigned to the Hernquist profile.

We fit our streams with cylinders whose axes lie in the direction of the stream and densities that decline with distance from the cylinder axis as a Gaussian.  Each stream fit requires six parameters: $\varepsilon$, the number of MSTO stars in the stream; $\mu$, the angular position of the stream center along the stripe; $R$, the distance of the stream center from the Sun; $\theta$ and $\phi$; the orientation of the cylinder axis; and $\sigma$, the standard deviation of the Gaussian density decline with distance from the cylinder axis.  We restrict the stream center to lie on the center of the stripe ($\nu = 0$) to prevent degeneracies in our parameter space.

\section{Fitting Results}

This model, with three streams per stripe, was fit to fourteen SDSS stripes in the north Galactic cap.  The rationale for fitting three streams was that we expected the largest overdensities in the spheroid to be the Sagittarius (Sgr) leading tail \citep{Majewski2003, Belokurov2006, newberg2007, Koposov2012, Belokurov2014, Hernitschek2017}; a fainter stream that appears to split off from the Sgr stream \citep[the ``bifurcated" stream][]{Belokurov2006}; and an overdensity in Virgo. The Virgo Overdentisy is a photometric overdensity which contains kinematic substructures like the Virgo Stellar Stream \citep{Vivas2001, Duffau2006, Juric2008} and other smaller moving groups \citep{newberg2007, duffau2014}. The stream parameters were constrained within the ranges: $-20<\varepsilon<20$, $2<R<100$ kpc, $0<\theta<3.14$ radians, $0<\phi<3.14$ radians, and $0.1<\sigma<25$ kpc.  $\mu$ was constrained to be with the stripe limits.

The smooth background fit parameters we obtained are given in Table \ref{ResultsTableHalo}.  They suggest an oblate stellar halo with a low disk fraction in our dataset; the low disk fraction is expected since the $(g-r)_0$ color cut was designed to select stars bluer than the thick disk turnoff.  The flattening parameter is surprisingly consistent; the stellar halo has a flattening of $0.58$ with a dispersion in the measurements of $0.04$, and is consistent with results previously reported by \cite{newby2013} even though we fit a slightly different background model.  The dispersion in measurements is somewhat larger than the statistical error in each measurement, which suggests systematic error due to an imperfect density model.  

\begin{figure}
\centering
\includegraphics[width=0.9\textwidth]{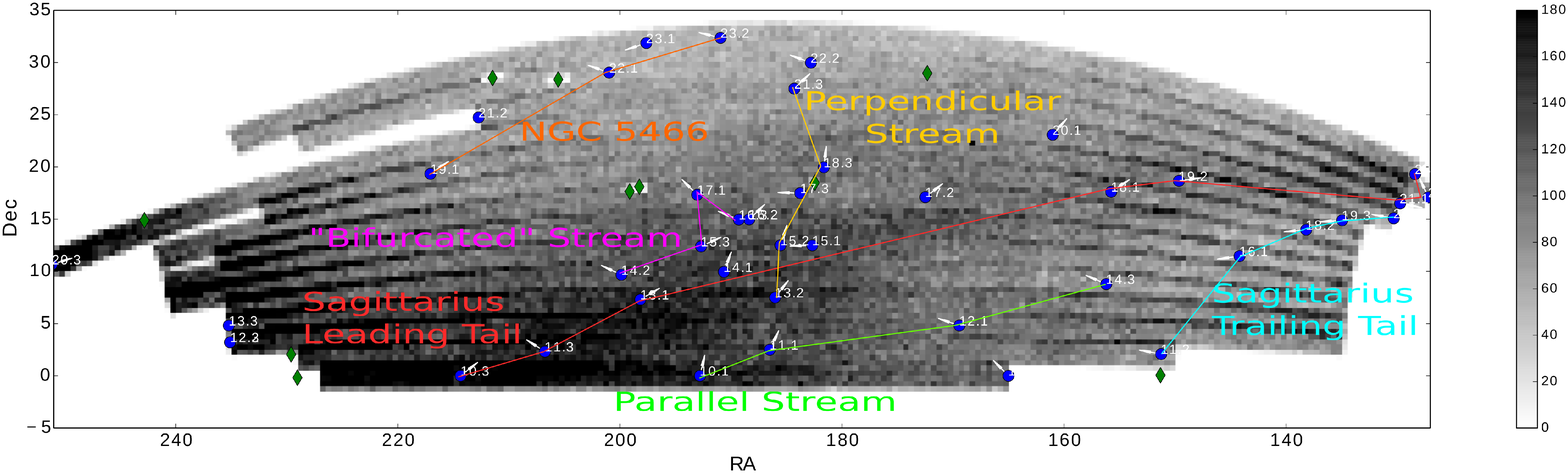}
\includegraphics[width=0.9\textwidth]{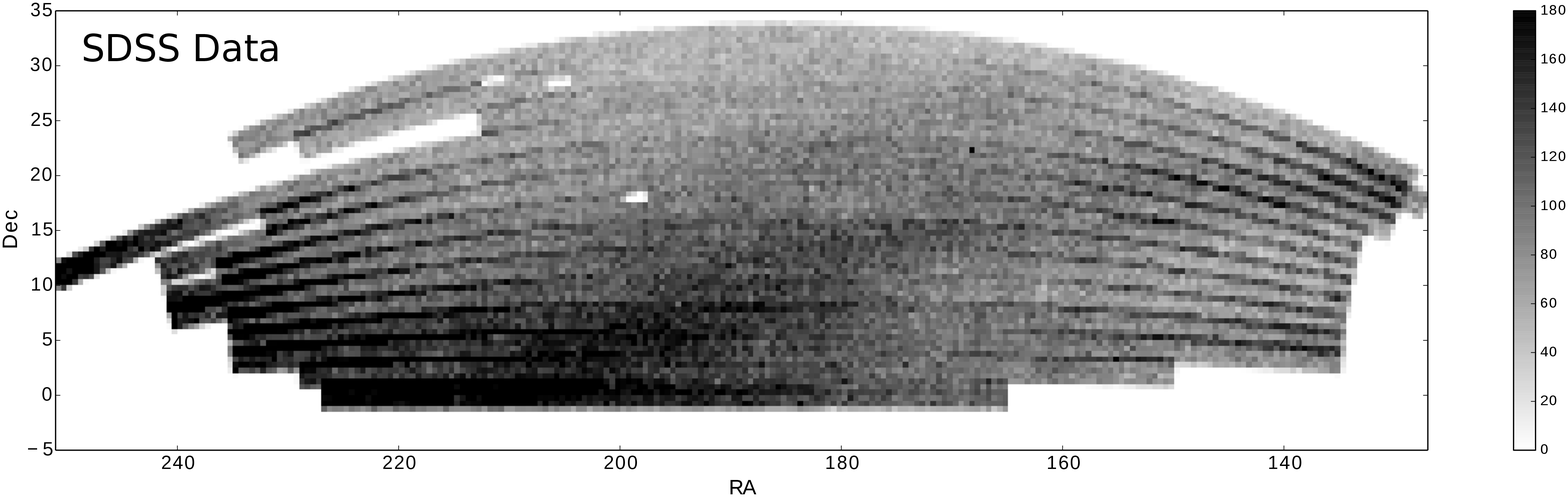}
\includegraphics[width=0.9\textwidth]{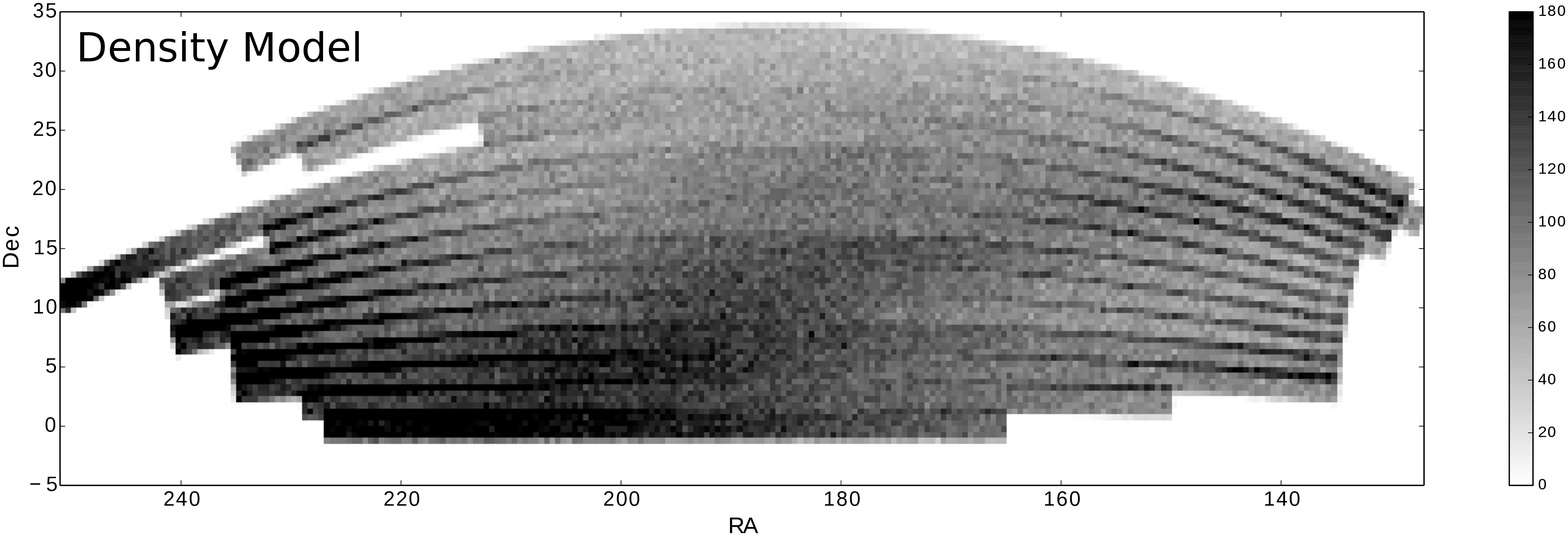}
\caption{RA and dec of streams centers (blue filled circles) fit by MilkyWay@home and comparison of density model to SDSS data.  Colored lines suggest the Sgr leading tail (red), the ``bifurcated" stream (magenta), the NGC 5466 stream (orange), the Perpendicular Stream (gold), the Sgr trailing tail (cyan), and the Parallel Stream (green).  Green diamonds show the positions of globular clusters.  The model density (lower panel) shows that we fit the SDSS data (middle panel) fairly well, but we are missing part of the ``bifurcated" stream.  }
\label{FieldofStreamsRemakeWithMarkings}
\end{figure}

\begin{figure}
\centering
\includegraphics[width=1.0\textwidth]{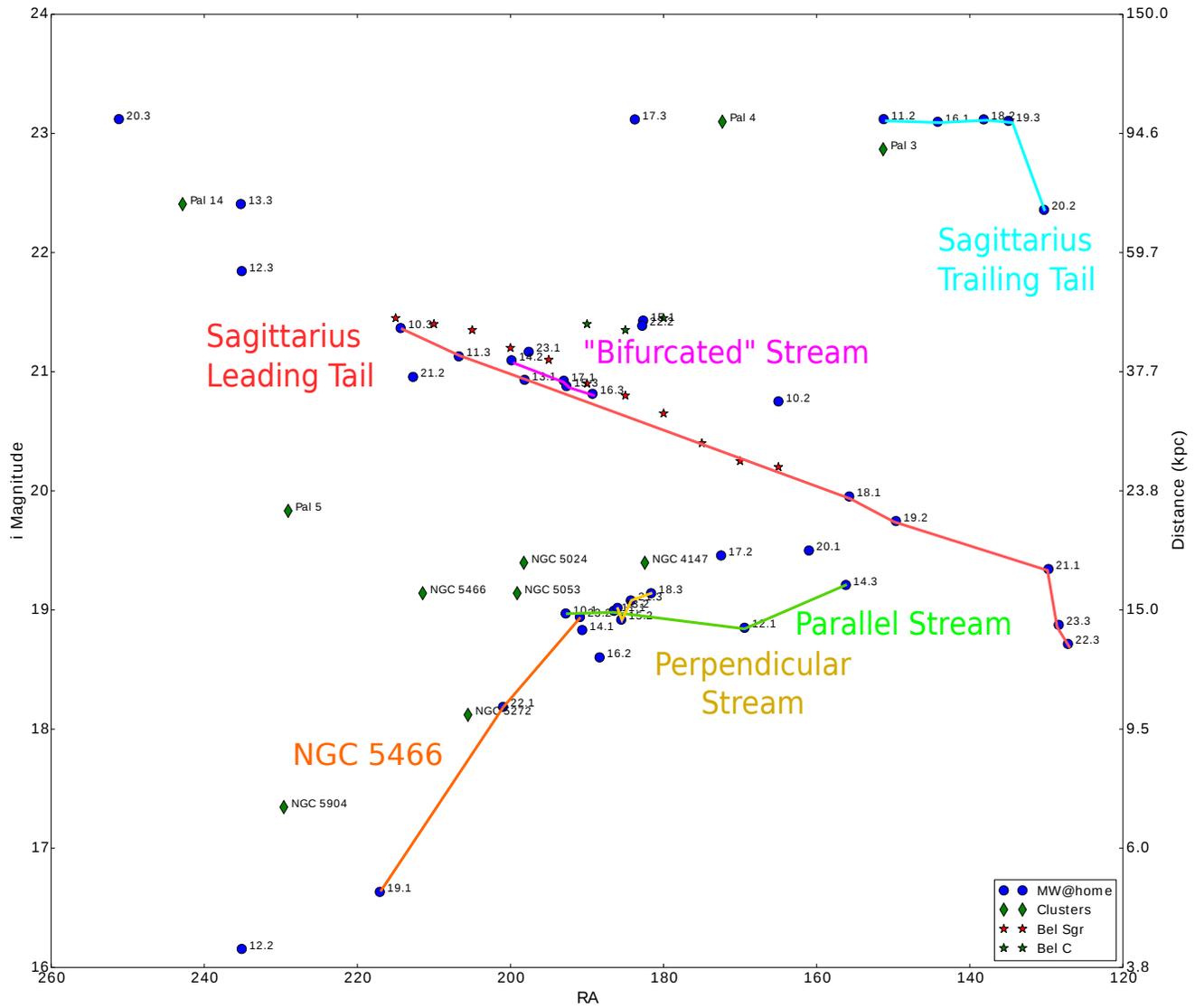}
\caption{RA and magnitude of stream centers fit by MilkyWay@home.  Symbols and stream colors are as described in Figure \ref{FieldofStreamsRemakeWithMarkings}.  A large number of stream centers are fit with $180<$ RA $<200$ and $10 <$ R $<18$ kpc. The Sgr stream positions from \cite{Belokurov2006} are shown with asterisks.}
\label{RADistanceRemakeWithMarkings}
\end{figure}

The positions of the 42 stream centers (three for each stripe) are shown in the top panel of Figure \ref{FieldofStreamsRemakeWithMarkings} and Figure \ref{RADistanceRemakeWithMarkings}.  Positions of known globular clusters in these figures are taken from the Harris catalog \citep{Harris1996}.  We compared these stream positions with the locations of known streams, and looked for alignments in position and distance.  

We identified seven candidate tidal stream fragments, as suggested in Figures \ref{FieldofStreamsRemakeWithMarkings} and \ref{RADistanceRemakeWithMarkings}.  The parameters for our stream fits can be found in Table \ref{ResultsTableStreams}.  The statistical errors are calculated from the width of the peak of the likelihood surface, using the method detailed in \cite{WeissSubmitted}.  Thirteen stream centers were not included in the table; half of these were discarded because they were located on the very end of a stripe, and the other half appeared to be placed between two or more identified streams.  

Four fragments appear to be associated with the Sgr stream: the leading tidal tail, the ``bifurcated" stream, Stream C from \cite{Belokurov2006}, and the trailing tail \citep{Belokurov2014}.  Note that the Sgr leading tail, that appears to dominate the sky in the density plots such as are shown in Figure \ref{FieldofStreamsRemakeWithMarkings}, is not identified in six of the fourteen stripes.  This is likely due to the presence of more than three streams in many of the stripes; in particular there are multiple streams that appear to cross in the center of the sky area, where the stellar density is highest.  Note that all of the ``bifurcated" stream identifications are in stripes that do not include Sgr stream identifications; all of these fits could be a combination of Sgr and the ``bifurcated" stream.  Also note that the Sgr trailing tail is located farther away than the majority of the stars in the sample, and in fact farther than we expected to fit substructure.  It is believed that the distant trailing tail connects to Stream C, but in this diagram that connection is not obvious due to the disparate sky positions of these two structures; the centers of the trailing tidal tail do not appear to be in the same plane as the leading tidal tail.

A third of our identified bits of substructure are at a distance of 15 kpc from the Sun, and most of them are inconsistent with the Sgr dwarf tidal stream.  Because they appear to separate into two roughly linear structures, we tentatively identify the Parallel Stream, approximately parallel to the Celestial Equator \citep[after][]{Sohn2016} and the Sgr stream, and the Perpendicular Stream, roughly perpendicular to the Celestial Equator.

We also note that three bits of substructure are aligned with the NGC 5466 stream \citep{Belokurov2006b, GrillmairandJohnson2006, Fellhauer2007}.  Note, however, that the identified substructure is much too wide to have a globular cluster progenitor.  Also, the apparent substructure is 2 kpc closer to the Sun in the direction of NGC 5466.  The apparently wide substructure, however, could easily enclose the globular cluster and its tidal stream.
  
\begin{figure}
\centering
\includegraphics[width=.7\textwidth]{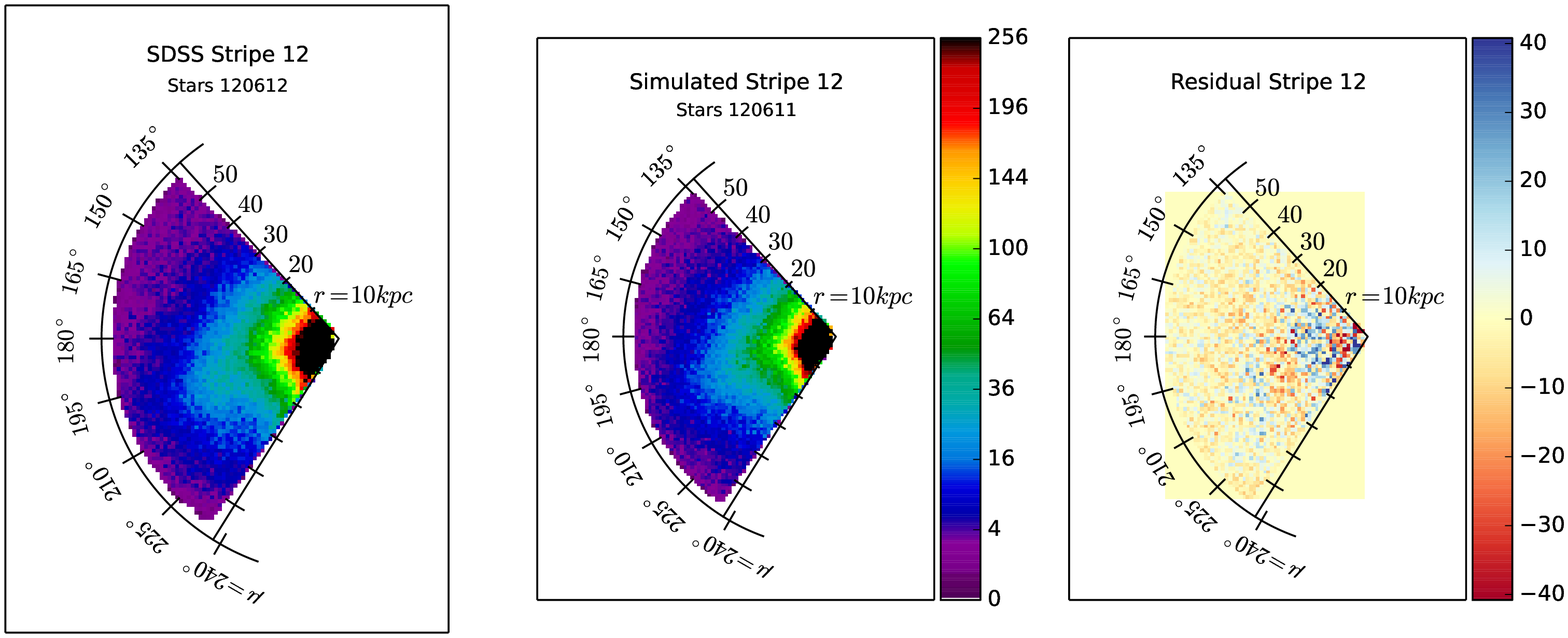}
\includegraphics[width=.7\textwidth]{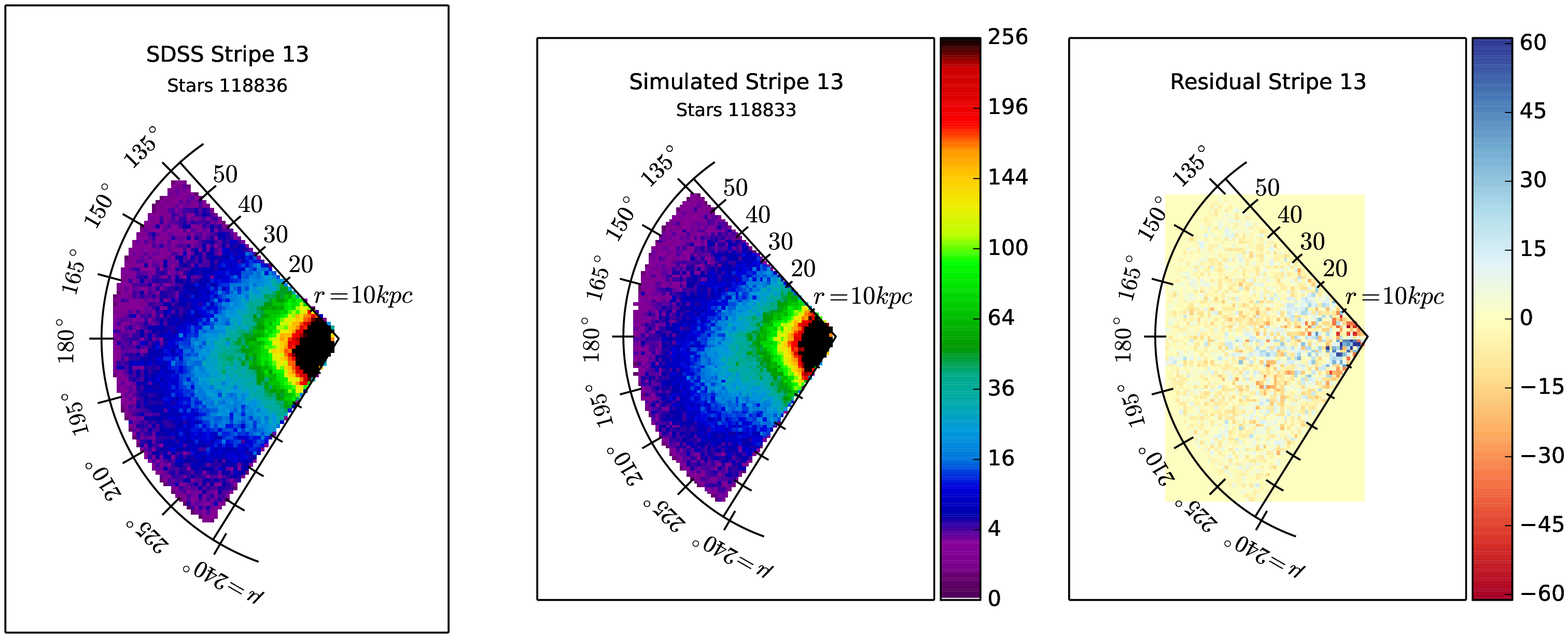}
\includegraphics[width=.7\textwidth]{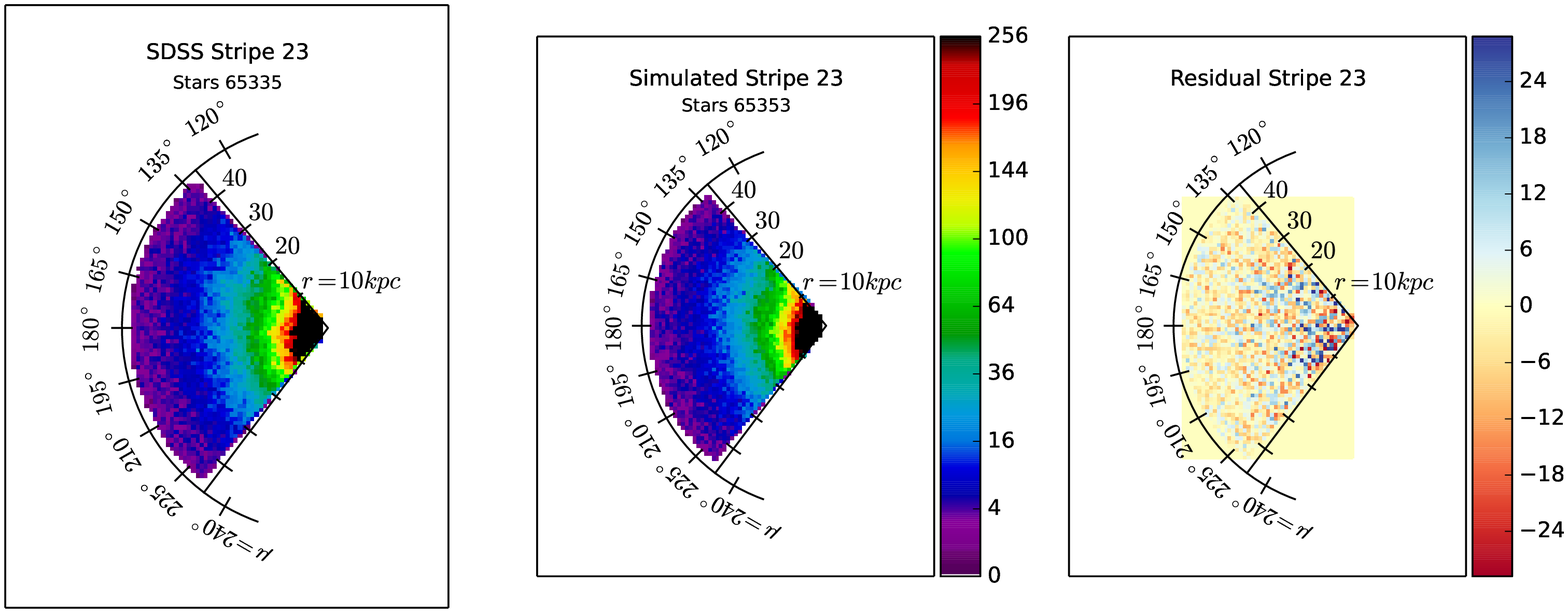}
\caption{Wedge plots and residuals for SDSS stripes 12,13, and 23, fit by MilkyWay@home.  Left panels show the SDSS data.  Middle panels show the density simulated from the fit model parameters, and right panels show the residuals between the two. Each pixel shows the MSTO density in a $1$ kpc by $1$ kpc by $2.5^{\circ}$ pixel.  Note that the color scale of the residual is different for each wedge.}
\label{FullSkyResiduals}
\end{figure}

A simulated dataset was created with the fit parameters for each stripe, using the process described in \citet{WeissSubmitted}.  A comparison of the density distribution on the sky of this generated data to the SDSS data is also shown in Figure \ref{FieldofStreamsRemakeWithMarkings} and the wedge plots in Figure \ref{FullSkyResiduals}.  A comparison of panels 2 and 3 in Figure \ref{FieldofStreamsRemakeWithMarkings} shows that we are missing some substructure in our model, in particular much of the ``bifurcated"  stream.  The incomplete characterization is confirmed by looking at the residual plots in the rightmost panels of Figure \ref{FullSkyResiduals}, in which we can see substructure missed in the optimization.

\section{Discussion}

Our results suggest the existence of far more than three streams in the volume explored. The middle stripes in principle could intersect all seven of the identified substructures, in addition to other unfit streams like the Orphan Stream.  The Orphan Stream is known to cross this region of the sky and is visible in the Field of Streams \citep{Belokurov2006}, but is not detected.  The stream can be clearly seen in a subset of our data with the apparent magnitude turnoff of stars in the Orphan Stream; it is possible that we are not fitting enough streams to detect it.  Our model can be extended to fit an arbitrary number of streams and future work will attempt to determine the optimal number of streams to fit to the data.

One of the effects of fitting too few streams is systematic uncertainty in our fit stream parameters, especially direction and width.  The width might be widened or the center shifted if the algorithm attempted to fit multiple halo substructures to one model stream.  However, since many of the stream centers fit known streams and many others are aligned, we are confident that we are tracing stellar halo substructure in the north Galactic cap.  The newly proposed streams connect 3-4 independently fit stream centers, further decreasing the probability that they follow spurious substructure.  Future work will explore the effects of fitting when the model is less complex than the actual stellar halo, in addition to adding more streams to the model.

Our Sgr leading tail candidate stream centers are consistent with those found in \cite{Belokurov2006}, \cite{newberg2007}, \cite{Belokurov2014}, and \cite{newby2013} for 8 out of 14 stripes.  Our model did not fit the leading tail in stripes 12, 14, 15, 16, 17 and 20, possibly because three streams were not sufficient to describe the substructure in these stripes for our model to fit it all with only three streams.  

Positions for the ``bifurcated" stream were found in four stripes.  The results are consistent in distance with those found in \cite{Belokurov2006}, \cite{newberg2007}, and \cite{Belokurov2014}, but the sky positions are inconsistent.  This is possibly due to an attempt to fit both the ``bifurcated" stream and the Sgr leading tail with a single stream in these stripes.

We find a group of stream centers at 100 kpc in stripes 11, 16, 18, 19, and 20 between (RA, dec) positions of $(130^{\circ}, 15^{\circ})$ and $(151^{\circ}, 2^{\circ})$.  These stream centers could be the Sgr trailing tail, since \cite{Belokurov2014} found this stream between 70 and 100 kpc in approximately this sky direction.  However, \cite{Belokurov2014} did not specify the stream's angular position on the sky, so the association is somewhat tentative.  It should be noted that our distance and $\mu$ uncertainties are large, and our measured stream widths are extremely large; large systematic uncertainty is likely because our distances are constrained to be 100 kpc or less, and several of the fit results are at this fitting limit.  This could indicate that we have a poor fit to the trailing stream.  Oddly, the width of the Stream C candidate, which could also be associated with the trailing tidal tail, is also extremely large.  But Stream C was also found on a stripe in which the Sgr dwarf tidal stream was not identified, so this structure could also be fitting two streams simultaneously and thus have an erroneous width.  However, if the large widths and precessed orbital plane are confirmed, these results will have implications for the flattening of the Milky Way gravitational potential; wide streams can form in an oblate or very lumpy gravitational potential \citep{Ibata2001, SiegalGaskinsValluri2008, Ngan2016, Sandford2017}. 

The proposed Parallel and Perpendicular streams cross each other in the same sky location and at the same distance as the previously identified overdensity in Virgo.  A clear explanation for the Virgo Overdensity and the associated Virgo Stellar Stream has been elusive and recent results suggest some or all of the overdensity could be possibly the result of an ancient massive merger that came in on a highly eccentric orbit, known as the Gaia ``Sausage" \citep{simion2018}.  \cite{CarlinVirgo} also attributed the Virgo Overdensity to an accretion event on a highly eccentric orbit that is aligned, within errors, with the Parallel Stream in our data, and also puffs out to include disrupted stars at the distance at which \cite{Sohn2016} found three stars with coherent proper motions in an HST pencil beam at (RA, dec) = ($158^{\circ}$, $7^{\circ}$).  The three stars lie between our Parallel Stream centers in stripes 12 and 14, at a distance of 32 kpc, which is twice our distance of 14-16.5 kpc for this stream.  A full explanation of the relationship between all of these structures is beyond the scope of this paper, but our preliminary results suggest there could be two substructures crossing in the highest density region of Virgo Overdensity, near (RA, dec) = ($191^{\circ}$, $-3^{\circ}$).  Our Perpendicular Stream extends from (RA, dec) = ($186^{\circ}$, $7.5^{\circ}$) to ($184^{\circ}$, $27.5^{\circ}$) at a heliocentric distance of 15 kpc.  If extrapolated $5^{\circ}$ to the south, it crosses the Parallel Stream precisely in this highest density region.  This could explain multiple line-of-sight velocities seen in this region by \cite{Duffau2006} and \cite{Newberg2009}.  The candidate Perpendicular Stream passes close to the globular cluster NGC 4147, which could be associated.

The candidate stellar stream near the globular cluster NGC 5466 extends from an (RA, dec) of $(217^{\circ}, 19^{\circ})$ to $(191^{\circ}, 32.3^{\circ})$, at a distance of 5 to 15 kpc.  This is close to, but about $5^\circ$ south of, the NGC 5466 stream as depicted in \citet{GrillmairCarlin2016}.  It is possible that our NGC 5466 stream could be associated with a dwarf galaxy that was the progenitor of this stream. This is the weakest stream detection, with only 3 potential stream centers. 

\section{Conclusions}

Using improved fitting methods outlined in \cite{WeissSubmitted}, we fit 7 candidate stream fragments in the north Galactic cap using blue MSTO stars from the SDSS as tracers:

\begin{enumerate}

\item The candidate Perpendicular Stream, if confirmed, would be a new discovery, and could contribute to the overdensity of stars found in Virgo.  The globular cluster NGC 4147 could be associated with this potential stream.

\item Another candidate stream, if confirmed, could connect the previously discovered Parallel Stream identified by \cite{Sohn2016} with the Virgo Overdensity.  

\item A third candidate stream is aligned with the previously identified NGC 5466 stream, but is offset by about $5^\circ$ and appears to be considerably wider and somewhat closer to the Sun.  It is possible that this stream progenitor was a dwarf galaxy that contained the globular cluster.  

\item Four additional candidate streams are associated with previously identified structures that are believed to arise from the tidal disruption of the Sagittarius dwarf galaxy; our stream centers for the Sgr structures (leading tail, ``bifurcated" stream, Stream C and the trailing tidal tail) are reasonably consistent with previous literature.  

\end{enumerate}

We have shown that there is significant halo density substructure beyond what was expected from a smooth halo plus a single Virgo Overdensity, the Sgr dwarf tidal stream and the associated ``bifurcated" stream. To fit halo substructure, our halo model must be more complex and include additional streams.

The smooth background component was consistently and independently fit in 14 stripes with an average flattening of $q=0.58$ and a measurement dispersion of 0.04.  The consistency of the smooth component is notable given the challenge encountered in fitting the stellar substructure.

\acknowledgements

We particularly thank the MilkyWay@home volunteers 
who have made this research possible. This work was supported by The Marvin Clan, Babette Josephs, Manit Limlamai, the 2015 Crowd Funding Campaign to Support Milky Way Research, NSF grant No. AST 16-15688 and the NASA/NY Space Grant.  Funding for the SDSS and SDSS-II has been provided by the Alfred P. Sloan Foundation, the Participating Institutions, the National Science Foundation, the U.S. Department of Energy, the National Aeronautics and Space Administration, the Japanese Monbukagakusho, the Max Planck Society, and the Higher Education Funding Council for England. The SDSS Web Site is http://www.sdss.org/.

\begin{table}
 \centering
 \caption{\label{ResultsTableHalo}MilkyWay@home smooth background parameters for a Hernquist halo with a thick disk.  These results are all show a low disk weight and an oblate halo.}
 \begin{tabular}{|c c c|} 
 \hline
 \multicolumn{3}{|c|}{Halo Fits from MilkyWay@home} \\
 \hline\hline 
 Stripe & $\varepsilon_{sph}$ & $q$ \\ [0.5ex] 
 \hline 
10 & $0.9979\pm0.0004$ & $0.54\pm0.02$ \\
\hline
11 & $0.9980\pm0.0003$ & $0.54\pm0.02$ \\
\hline
12 & $0.9972\pm0.0003$ & $0.61\pm0.01$ \\
\hline
13 & $0.9982\pm0.0003$ & $0.54\pm0.01$ \\
\hline
14 & $0.9972\pm0.0003$ & $0.62\pm0.01$ \\
\hline
15 & $0.9976\pm0.0003$ & $0.57\pm0.01$ \\
\hline
16 & $0.9970\pm0.0004$ & $0.58\pm0.02$ \\
\hline
17 & $0.9968\pm0.0003$ & $0.62\pm0.02$ \\
\hline
18 & $0.9969\pm0.0004$ & $0.62\pm0.01$ \\
\hline
19 & $0.9968\pm0.0005$ & $0.57\pm0.02$ \\
\hline
20 & $0.9967\pm0.0003$ & $0.64\pm0.02$ \\
\hline
21 & $0.9977\pm0.0004$ & $0.62\pm0.03$ \\
\hline
22 & $0.9970\pm0.0007$ & $0.51\pm0.03$ \\
\hline
23 & $0.9979\pm0.0004$ & $0.55\pm0.03$ \\
\hline
\end{tabular}
\end{table}

\begin{table}
 \centering
 \caption{\label{ResultsTableStreams}Stream centers fit by MilkyWay@home, organized by suggested substructure.  Stream centers that don't have any clear relation to known substructure or other stream centers were not included.  The ``Stripe" column in this table corresponds to stream center labels in Figures \ref{FieldofStreamsRemakeWithMarkings} and \ref{RADistanceRemakeWithMarkings}.}
 \begin{tabular}{|l c c c c c c c c|} 
 \hline
  \multicolumn{9}{|c|}{Sagittarius} \\
 \hline
 Stripe & $\varepsilon$ & $l$ (deg) & $b$ (deg) & $\mu$ (deg) & R (kpc) & $\theta$ (rad) & $\phi$ (rad) & $\sigma$ (kpc) \\
 \hline
10.3 & $-0.53\pm0.07$ & $343.7$ & $55.9$ & $214.4\pm0.4$ & $44.6\pm0.3$ & $1.1\pm0.05$ & $3.14\pm0.07$ & $7.1\pm0.5$ \\
\hline
11.3 & $-1.46\pm0.09$ & $333.5$ & $61.9$ & $206.7\pm0.5$ & $40.0\pm0.6$ & $1.17\pm0.09$ & $1.03\pm0.08$ & $4.7\pm0.6$ \\
\hline
13.1 & $-1.45\pm0.10$ & $318.1$ & $69.6$ & $198.0\pm1.4$ & $36.5\pm1.1$ & $1.59\pm0.06$ & $3.14\pm0.23$ & $4.9\pm1.7$ \\
\hline
18.1 & $-2.04\pm0.15$ & $220.6$ & $54.5$ & $157.2\pm0.6$ & $23.3\pm0.4$ & $2.23\pm0.13$ & $2.44\pm0.26$ & $2.5\pm0.6$ \\
\hline
19.2 & $-1.93\pm0.08$ & $215.6$ & $49.5$ & $151.8\pm0.4$ & $21.1\pm0.3$ & $2.57\pm0.10$ & $2.79\pm0.18$ & $1.0\pm0.2$ \\
\hline
21.1 & $-1.51\pm0.10$ & $209.2$ & $31.0$ & $133.0\pm9.1$ & $17.6\pm0.9$ & $2.78\pm0.05$ & $3.14\pm0.12$ & $2.2\pm0.5$ \\
\hline
22.3 & $-2.38\pm0.14$ & $207.5$ & $29.0$ & $131.0\pm4.8$ & $13.2\pm2.1$ & $0.75\pm0.11$ & $3.07\pm0.21$ & $1.6\pm0.3$ \\
\hline
23.3 & $-3.00\pm0.14$ & $205.6$ & $30.8$ & $133.0\pm2.4$ & $14.2\pm1.1$ & $2.25\pm0.13$ & $1.11\pm0.31$ & $1.7\pm0.4$ \\
\hline\hline 
\multicolumn{9}{|c|}{``Bifurcated" Stream} \\
\hline
14.2 & $-1.01\pm0.06$ & $325.0$ & $71.3$ & $199.6\pm1.2$ & $39.4\pm0.9$ & $1.12\pm0.04$ & $0.93\pm0.04$ & $8.9\pm0.6$ \\
\hline
15.3 & $-1.73\pm0.11$ & $302.3$ & $75.3$ & $192.5\pm1.8$ & $35.6\pm1.1$ & $1.94\pm0.05$ & $3.14\pm0.09$ & $4.5\pm0.8$ \\
\hline
16.3 & $-1.77\pm0.16$ & $286.9$ & $77.4$ & $189.1\pm1.9$ & $34.6\pm1.0$ & $1.12\pm0.08$ & $0.9\pm0.09$ & $3.9\pm0.8$ \\
\hline
17.1 & $-1.19\pm0.14$ & $303.9$ & $80.2$ & $192.7\pm1.5$ & $36.4\pm1.2$ & $1.56\pm0.13$ & $1.43\pm0.19$ & $10.1\pm1.0$ \\
\hline\hline
\multicolumn{9}{|c|}{Sagittarius Trailing Tail} \\
\hline
11.2 & $-0.83\pm0.09$ & $237.9$ & $43.0$ & $151.3\pm10.3$ & $100.0\pm27.6$ & $1.85\pm0.09$ & $0.8\pm0.08$ & $25.0\pm4.6$ \\
 \hline
16.1 & $-1.39\pm0.1$ & $222.1$ & $41.8$ & $145.2\pm4.4$ & $99.0\pm29.2$ & $1.9\pm0.14$ & $0.46\pm0.09$ & $21.0\pm4.5$ \\
\hline
18.2 & $-1.40\pm0.09$ & $215.9$ & $37.6$ & $140.0\pm4.9$ & $99.9\pm14.4$ & $1.86\pm0.08$ & $0.42\pm0.03$ & $16.1\pm2.2$ \\
\hline
19.3 & $-1.00\pm0.10$ & $213.3$ & $35.0$ & $137.2\pm3.6$ & $99.3\pm17.7$ & $1.81\pm0.12$ & $0.39\pm0.04$ & $18.4\pm3.1$ \\
\hline
20.2 & $-2.35\pm0.39$ & $211.0$ & $31.0$ & $133.0\pm6.3$ & $70.4\pm3.2$ & $1.46\pm0.12$ & $0.42\pm0.02$ & $7.3\pm2.1$ \\
\hline\hline 
\multicolumn{9}{|c|}{Parallel Stream} \\
\hline
10.1 & $-0.89\pm0.17$ & $302.8$ & $62.9$ & $192.8\pm1.4$ & $14.8\pm0.9$ & $0.39\pm0.12$ & $2.57\pm0.26$ & $4.5\pm0.4$ \\
\hline
11.1 & $-0.79\pm0.09$ & $288.0$ & $64.6$ & $186.5\pm1.1$ & $14.9\pm0.6$ & $0.82\pm0.15$ & $2.74\pm0.16$ & $5.2\pm0.3$ \\
\hline
12.1 & $-0.69\pm0.06$ & $253.7$ & $58.4$ & $169.5\pm7.1$ & $14.0\pm1.8$ & $0.43\pm0.02$ & $0.28\pm0.04$ & $6.8\pm0.3$ \\
\hline
14.3 & $-3.07\pm0.18$ & $233.9$ & $50.8$ & $156.6\pm2.7$ & $16.5\pm0.7$ & $0.33\pm0.14$ & $0.59\pm0.46$ & $1.4\pm0.7$ \\
\hline\hline 
\multicolumn{9}{|c|}{Perpendicular Stream} \\
 \hline
13.2 & $-1.12\pm0.13$ & $283.4$ & $69.3$ & $186.0\pm0.9$ & $15.1\pm0.5$ & $1.18\pm0.16$ & $2.82\pm0.22$ & $4.8\pm0.8$ \\
\hline
15.2 & $-1.03\pm0.10$ & $276.4$ & $73.8$ & $185.5\pm1.1$ & $14.4\pm1.0$ & $0.55\pm0.14$ & $2.68\pm0.14$ & $5.2\pm0.2$ \\
\hline
18.3 & $-0.93\pm0.07$ & $245.4$ & $77.5$ & $181.8\pm1.6$ & $16.0\pm2.1$ & $2.61\pm0.05$ & $1.83\pm0.07$ & $5.8\pm0.3$ \\
\hline
21.3 & $-1.71\pm0.20$ & $208.2$ & $82.4$ & $184.4\pm1.4$ & $15.6\pm0.5$ & $1.34\pm0.24$ & $2.8\pm0.18$ & $3.0\pm0.6$ \\
\hline\hline 
\multicolumn{9}{|c|}{NGC 5466 Stream} \\ 
 \hline
19.1 & $-0.36\pm0.07$ & $18.8$ & $66.5$ & $215.1\pm12.4$ & $5.0\pm0.4$ & $0.49\pm0.04$ & $3.02\pm0.03$ & $6.2\pm0.3$ \\
\hline
22.1 & $-0.23\pm0.07$ & $49.7$ & $82.6$ & $198.9\pm3.4$ & $10.3\pm0.5$ & $2.09\pm0.04$ & $1.08\pm0.04$ & $6.2\pm0.4$ \\
\hline
23.2 & $-1.17\pm0.16$ & $140.2$ & $84.5$ & $190.0\pm5.7$ & $14.6\pm0.7$ & $1.77\pm0.06$ & $0.86\pm0.05$ & $3.7\pm0.5$ \\
\hline\hline
\multicolumn{9}{|c|}{Stream C \citep{Belokurov2006}} \\ 
 \hline
15.1 & $-1.42\pm0.09$ & $267.8$ & $72.5$ & $182.7\pm14.2$ & $45.9\pm7.3$ & $1.91\pm0.13$ & $0.60\pm0.20$ & $25.0\pm4.9$ \\
\hline
\end{tabular}
\end{table}

\newpage
\bibliographystyle{apj}
\bibliography{references.bib}

\newcommand{\noop}[1]{}
\begin{thebibliography}{}
\expandafter\ifx\csname natexlab\endcsname\relax\def\natexlab#1{#1}\fi

\bibitem[{{Abazajian} {et~al.}(2009){Abazajian}, {Adelman-McCarthy},
  {Ag{\"u}eros}, {Allam}, {Allende Prieto}, {An}, {Anderson}, {Anderson},
  {Annis}, {Bahcall}, \& et~al.}]{DR7Abazajian2009}
{Abazajian}, K.~N., {Adelman-McCarthy}, J.~K., {Ag{\"u}eros}, M.~A., {et~al.}
  2009, \apjs, 182, 543

\bibitem[{{Belokurov} {et~al.}(2006){Belokurov}, {Evans}, {Irwin}, {Hewett}, \&
  {Wilkinson}}]{Belokurov2006b}
{Belokurov}, V., {Evans}, N.~W., {Irwin}, M.~J., {Hewett}, P.~C., \&
  {Wilkinson}, M.~I. 2006, \apjl, 637, L29

\bibitem[{Belokurov {et~al.}(2006)Belokurov, Zucker, Evans, Gilmore, Vidrih,
  Bramich, Newberg, Wyse, Irwin, Fellhauer, Hewett, Walton, Wilkinson, Cole,
  Yanny, Rockosi, Beers, Bell, Brinkmann, \v{Z}. Ivezi\'{c}, \&
  Lupton}]{Belokurov2006}
Belokurov, V., Zucker, D.~B., Evans, N.~W., {et~al.} 2006, \apjl, 642, L137

\bibitem[{{Belokurov} {et~al.}(2014){Belokurov}, {Koposov}, {Evans},
  {Pe{\~n}arrubia}, {Irwin}, {Smith}, {Lewis}, {Gieles}, {Wilkinson},
  {Gilmore}, {Olszewski}, \& {Niederste-Ostholt}}]{Belokurov2014}
{Belokurov}, V., {Koposov}, S.~E., {Evans}, N.~W., {et~al.} 2014, \mnras, 437,
  116

\bibitem[{Carlin {et~al.}(2012)Carlin, Yam, Casetti-Dinescu, Willett, Newberg,
  Majewski, \& Girard}]{CarlinVirgo}
Carlin, J.~L., Yam, W., Casetti-Dinescu, D.~I., {et~al.} 2012, \apj, 753, 145

\bibitem[{Cole {et~al.}(2008)Cole, Newberg, Magdon-Ismail, Desell, Dawsey,
  Hayashi, Liu, Purnell, Szymanski, Varela, Willett, \& Wisniewski}]{cole2008}
Cole, N., Newberg, H.~J., Magdon-Ismail, M., {et~al.} 2008, \apj, 683, 750

\bibitem[{{Desell} {et~al.}(2007){Desell}, Cole, Magdon-Ismail, Newberg,
  Szymanski, \& Varela}]{desell2007}
{Desell}, T., Cole, N., Magdon-Ismail, M., {et~al.} 2007, in IEEE International
  Conference on e-Science and Grid Computing (Bangalore, India: IEEE), 337--344

\bibitem[{{Desell} {et~al.}(2011){Desell}, Magdon-Ismail, Newberg, Newberg,
  Szymanski, \& Varela}]{desell2011}
{Desell}, T., Magdon-Ismail, M., Newberg, H., {et~al.} 2011, in International
  Conference on Computational Intelligence and Software Engineering (CiSE),
  Wuhan, China

\bibitem[{{Desell} {et~al.}(2017){Desell}, {Magdon-Ismail}, {Newberg},
  {Newberg}, {Szymanski}, \& {Varela}}]{desell2017}
{Desell}, T., {Magdon-Ismail}, M., {Newberg}, H., {et~al.} 2017,
  arXiv:1702.02204

\bibitem[{{Desell} {et~al.}(2010){Desell}, Magdon-Ismail, Szymanski, Varela,
  Newberg, \& Anderson}]{desell2010}
{Desell}, T., Magdon-Ismail, M., Szymanski, B., {et~al.} 2010, in Proceedings
  of the 10th International Conference on Distributed Applications and
  Interoperable Systems, Amsterdam, Netherlands, 29

\bibitem[{{Duffau} {et~al.}(2014){Duffau}, {Vivas}, {Zinn}, {M{\'e}ndez}, \&
  {Ruiz}}]{duffau2014}
{Duffau}, S., {Vivas}, A.~K., {Zinn}, R., {M{\'e}ndez}, R.~A., \& {Ruiz}, M.~T.
  2014, \aap, 566, A118

\bibitem[{{Duffau} {et~al.}(2006){Duffau}, {Zinn}, {Vivas}, {Carraro},
  {M{\'e}ndez}, {Winnick}, \& {Gallart}}]{Duffau2006}
{Duffau}, S., {Zinn}, R., {Vivas}, A.~K., {et~al.} 2006, \apjl, 636, L97

\bibitem[{{Fellhauer} {et~al.}(2007){Fellhauer}, {Evans}, {Belokurov},
  {Wilkinson}, \& {Gilmore}}]{Fellhauer2007}
{Fellhauer}, M., {Evans}, N.~W., {Belokurov}, V., {Wilkinson}, M.~I., \&
  {Gilmore}, G. 2007, \mnras, 380, 749

\bibitem[{{Grillmair} \& {Carlin}(2016)}]{GrillmairCarlin2016}
{Grillmair}, C.~J., \& {Carlin}, J.~L. 2016, in Astrophysics and Space Science
  Library, Vol. 420, Tidal Streams in the Local Group and Beyond, ed. H.~J.
  {Newberg} \& J.~L. {Carlin} (Cambridge, MA: Springer International
  Publishing), 87

\bibitem[{{Grillmair} \& {Johnson}(2006)}]{GrillmairandJohnson2006}
{Grillmair}, C.~J., \& {Johnson}, R. 2006, \apjl, 639, L17

\bibitem[{{Harris}(1996)}]{Harris1996}
{Harris}, W.~E. 1996, \aj, 112, 1487

\bibitem[{{Hernitschek} {et~al.}(2017){Hernitschek}, {Sesar}, {Rix},
  {Belokurov}, {Martinez-Delgado}, {Martin}, {Kaiser}, {Hodapp}, {Chambers},
  {Wainscoat}, {Magnier}, {Kudritzki}, {Metcalfe}, \&
  {Draper}}]{Hernitschek2017}
{Hernitschek}, N., {Sesar}, B., {Rix}, H.-W., {et~al.} 2017, \apj, 850, 96

\bibitem[{Hernquist(1990)}]{hernquist}
Hernquist, L. 1990, \apj, 356, 359

\bibitem[{{Ibata} {et~al.}(2001){Ibata}, {Lewis}, {Irwin}, {Totten}, \&
  {Quinn}}]{Ibata2001}
{Ibata}, R., {Lewis}, G.~F., {Irwin}, M., {Totten}, E., \& {Quinn}, T. 2001,
  \apj, 551, 294

\bibitem[{{Juri{\'c}} {et~al.}(2008){Juri{\'c}}, {Ivezi{\'c}}, {Brooks},
  {Lupton}, {Schlegel}, {Finkbeiner}, {Padmanabhan}, {Bond}, {Sesar},
  {Rockosi}, {Knapp}, {Gunn}, {Sumi}, {Schneider}, {Barentine}, {Brewington},
  {Brinkmann}, {Fukugita}, {Harvanek}, {Kleinman}, {Krzesinski}, {Long},
  {Neilsen}, {Nitta}, {Snedden}, \& {York}}]{Juric2008}
{Juri{\'c}}, M., {Ivezi{\'c}}, {\v Z}., {Brooks}, A., {et~al.} 2008, \apj, 673,
  864

\bibitem[{{Koposov} {et~al.}(2012){Koposov}, {Belokurov}, {Evans}, {Gilmore},
  {Gieles}, {Irwin}, {Lewis}, {Niederste-Ostholt}, {Pe{\~n}arrubia}, {Smith},
  {Bizyaev}, {Malanushenko}, {Malanushenko}, {Schneider}, \&
  {Wyse}}]{Koposov2012}
{Koposov}, S.~E., {Belokurov}, V., {Evans}, N.~W., {et~al.} 2012, \apj, 750, 80

\bibitem[{{Majewski} {et~al.}(2003){Majewski}, {Skrutskie}, {Weinberg}, \&
  {Ostheimer}}]{Majewski2003}
{Majewski}, S.~R., {Skrutskie}, M.~F., {Weinberg}, M.~D., \& {Ostheimer}, J.~C.
  2003, \apj, 599, 1082

\bibitem[{Newberg(2012)}]{newberg2013}
Newberg, H.~J. 2012, Proceedings of the International Astronomical Union, 8, 74

\bibitem[{{Newberg} {et~al.}(2014){Newberg}, {Newby}, {Desell},
  {Magdon-Ismail}, {Szymanski}, \& {Varela}}]{Newberg2014}
{Newberg}, H.~J., {Newby}, M., {Desell}, T., {et~al.} 2014, in IAU Symposium,
  Vol. 298, Setting the scene for Gaia and LAMOST, ed. S.~{Feltzing},
  G.~{Zhao}, N.~A. {Walton}, \& P.~{Whitelock}, 98--104

\bibitem[{Newberg \& Yanny(2006)}]{NewbergYanny2006}
Newberg, H.~J., \& Yanny, B. 2006, JPhCS, 47, 195

\bibitem[{{Newberg} {et~al.}(2007){Newberg}, {Yanny}, {Cole}, {Beers}, {Re
  Fiorentin}, {Schneider}, \& {Wilhelm}}]{newberg2007}
{Newberg}, H.~J., {Yanny}, B., {Cole}, N., {et~al.} 2007, \apj, 668, 221

\bibitem[{{Newberg} {et~al.}(2009){Newberg}, {Yanny}, \&
  {Willett}}]{Newberg2009}
{Newberg}, H.~J., {Yanny}, B., \& {Willett}, B.~A. 2009, \apjl, 700, L61

\bibitem[{Newberg {et~al.}(2002)Newberg, Yanny, Rockosi, Grebel, Rix,
  Brinkmann, Istvan, Csabai, Hennessy, Hindsley, Ibata, \v{Z}eljko Ivezi\'{c},
  Lamb, Nash, Michael, Odenkirchen, Rave, Schneider, Smith, Stolte, \&
  York}]{newberg2002}
Newberg, H.~J., Yanny, B., Rockosi, C., {et~al.} 2002, \apj, 569, 245

\bibitem[{Newby {et~al.}(2011)Newby, Newberg, Simones, Cole, \&
  Monaco}]{newby2011}
Newby, M., Newberg, H.~J., Simones, J., Cole, N., \& Monaco, M. 2011, \apj,
  743, 187

\bibitem[{Newby {et~al.}(2013)Newby, Cole, Newberg, Desell, Magdon-Ismail,
  Szymanski, Varela, Willett, \& Yanny}]{newby2013}
Newby, M., Cole, N., Newberg, H.~J., {et~al.} 2013, \aj, 145, 163

\bibitem[{{Ngan} {et~al.}(2016){Ngan}, {Carlberg}, {Bozek}, {Wyse}, {Szalay},
  \& {Madau}}]{Ngan2016}
{Ngan}, W., {Carlberg}, R.~G., {Bozek}, B., {et~al.} 2016, \apj, 818, 194

\bibitem[{{Sandford} {et~al.}(2017){Sandford}, {K{\"u}pper}, {Johnston}, \&
  {Diemand}}]{Sandford2017}
{Sandford}, E., {K{\"u}pper}, A.~H.~W., {Johnston}, K.~V., \& {Diemand}, J.
  2017, \mnras, 470, 522

\bibitem[{{Schlegel} {et~al.}(1998){Schlegel}, {Finkbeiner}, \&
  {Davis}}]{SFD1998}
{Schlegel}, D.~J., {Finkbeiner}, D.~P., \& {Davis}, M. 1998, \apj, 500, 525

\bibitem[{{Siegal-Gaskins} \& {Valluri}(2008)}]{SiegalGaskinsValluri2008}
{Siegal-Gaskins}, J.~M., \& {Valluri}, M. 2008, \apj, 681, 40

\bibitem[{{Simion} {et~al.}(2018){Simion}, {Belokurov}, \&
  {Koposov}}]{simion2018}
{Simion}, I.~T., {Belokurov}, V., \& {Koposov}, S.~E. 2018, arXiv:1807.01335

\bibitem[{{Sohn} {et~al.}(2016){Sohn}, {van der Marel}, {Kallivayalil},
  {Majewski}, {Besla}, {Carlin}, {Law}, {Siegel}, \& {Anderson}}]{Sohn2016}
{Sohn}, S.~T., {van der Marel}, R.~P., {Kallivayalil}, N., {et~al.} 2016, \apj,
  833, 235

\bibitem[{Vivas {et~al.}(2001)Vivas, Zinn, Andrews, Bailyn, Baltay, Coppi,
  Ellman, Girard, Rabinowitz, Schaefer, Shin, Snyder, Sofia, van Altena, Abad,
  Bongiovanni, Briceño, Bruzual, Prugna, Herrera, Magris, Mateu, Pacheco,
  Sánchez, Sánchez, Schenner, Stock, Vicente, Vieira, Ferrín, Hernandez,
  Gebhard, Honeycutt, Mufson, Musser, \& Rengstorf}]{Vivas2001}
Vivas, A.~K., Zinn, R., Andrews, P., {et~al.} 2001, \apjl, 554, L33

\bibitem[{{Weiss} {et~al.}(2018){Weiss}, {Newberg}, {Newby}, \&
  {Desell}}]{WeissSubmitted}
{Weiss}, J., {Newberg}, H.~J., {Newby}, M., \& {Desell}, T. 2018, \apjs, 238,
  17

\bibitem[{York {et~al.}(2000)York, Adelman, John E.~Anderson, Anderson, Annis,
  Bahcall, Bakken, Barkhouser, Bastian, Berman, Boroski, Bracker, Briegel,
  Briggs, Brinkmann, Brunner, Burles, Carey, Carr, Castander, Chen, Colestock,
  Connolly, Crocker, Csabai, Czarapata, Davis, Doi, Dombeck, Eisenstein,
  Ellman, Elms, Evans, Fan, Federwitz, Fiscelli, Friedman, Frieman, Fukugita,
  Gillespie, Gunn, Gurbani, de~Haas, Haldeman, Harris, Hayes, Heckman,
  Hennessy, Hindsley, Holm, Holmgren, hao Huang, Hull, Husby, Ichikawa,
  Ichikawa, \v{Z}eljko Ivezi\'{c}, Kent, Kim, Kinney, Klaene, Kleinman,
  Kleinman, Knapp, Korienek, Kron, Kunszt, Lamb, Lee, Leger, Limmongkol,
  Lindenmeyer, Long, Loomis, Loveday, Lucinio, Lupton, MacKinnon, Mannery,
  Mantsch, Margon, McGehee, McKay, Meiksin, Merelli, Monet, Munn, Narayanan,
  Nash, Neilsen, Neswold, Newberg, Nichol, Nicinski, Nonino, Okada, Okamura,
  Ostriker, Owen, Pauls, Peoples, Peterson, Petravick, Pier, Pope, Pordes,
  Prosapio, Rechenmacher, Quinn, Richards, Richmond, Rivetta, Rockosi,
  Ruthmansdorfer, Sandford, Schlegel, Schneider, Sekiguchi, Sergey, Shimasaku,
  Siegmund, Smee, Smith, Snedden, Stone, Stoughton, Strauss, Stubbs, SubbaRao,
  Szalay, Szapudi, Szokoly, Thakar, Tremonti, Tucker, Uomoto, Berk, Vogeley,
  Waddell, i~Wang, Watanabe, Weinberg, Yanny, \& Yasuda}]{SDSSYork}
York, D.~G., Adelman, J., John E.~Anderson, J., {et~al.} 2000, \aj, 120, 1579

\end{thebibliography}
\end{document}